\documentclass{webofc}
\usepackage[varg]{txfonts}   
\usepackage{amsmath, bbm}
\usepackage[figuresright]{rotating}
\usepackage{capt-of}
\usepackage{pdfpages}
\usepackage{graphicx}

\newcommand{\La}{\Lambda}
\newcommand{\Si}{\Sigma}

\newcommand{\be}{\begin{eqnarray}}
\newcommand{\ee}{\end{eqnarray}}

\newlength{\feynwidth} 
\newlength{\feynwidthbig} 

\usepackage{xcolor} 

\begin{document}
\title{Status of the hyperon-nucleon interaction in chiral effective field theory
}
%
%

\author{\firstname{Johann} \lastname{Haidenbauer}\inst{1}\fnsep\thanks{\email{j.haidenbauer@fz-juelich.de}} 
   \and \firstname{Ulf-G.} \lastname{Mei{\ss}ner}\inst{2,1}\fnsep\thanks{\email{meissner@hiskp.uni-bonn.de}}
}

\institute{
Institute for Advanced Simulation, Institut f\"ur Kernphysik and J\"ulich Center for Hadron Physics,
Forschungszentrum J{\"u}lich, D-52425 J{\"u}lich, Germany 
\and
Helmholtz-Institut f\"ur Strahlen- und Kernphysik and Bethe Center for Theoretical Physics,
  Universit\"at Bonn, D-53115 Bonn, Germany
          }

\abstract{ 
The J\"ulich-Bonn group aims at an extensive study of the baryon-baryon ($BB$) interaction involving 
strange baryons ($\Lambda$, $\Sigma$, $\Xi$) within SU(3) chiral effective field theory. 
An overview of achievements and new developments over the past few years is provided. 
The topics covered are:
1)~Derivation of the leading charge-symmetry breaking (CSB) interaction for 
the $\Lambda N$ system and its application in a study of CSB effects in $A$=$4$ $\Lambda$-hypernuclei.
2)~Updated results for the $\Xi N$ interaction at NLO and predictions for $\Xi^- p$ correlation functions.
3)~Extension of the $\Lambda N$-$\Sigma N$ interaction to next-to-next-to-leading order.
}
\maketitle

\section{Introduction}

The J\"ulich-Bonn group explores the baryon-baryon ($BB$) interaction 
involving hyperons within SU(3) chiral effective field theory (EFT).
In this approach a potential is established via an expansion in terms of small 
momenta, subject to an appropriate power counting, so that the results can be 
improved systematically by going to higher orders,
while at the same time theoretical uncertainties can be estimated 
\cite{Epelbaum:2009,Machleidt:2011}. 
Furthermore, two- and three-baryon forces can be constructed in a consistent way.
The resulting interaction potentials can be readily employed in
standard two- and few-body calculations. They consist of contributions
from an increasing number of pseudoscalar-meson exchanges, determined
by the underlying chiral symmetry,
and of contact terms which encode the unresolved short-distance dynamics and whose 
strengths are parameterized by a priori unknown low-energy constants (LECs).

The studies, performed so far up to next-to-leading order (NLO) in the chiral
expansion, have shown that for the strangeness $S=-1$ ($\Lambda N$, $\Sigma N$)
\cite{YN2013,YN2019} and $S=-2$ ($\Lambda \Lambda$, $\Xi N$) \cite{YY2015,YY2019}
sectors a consistent and satisfactory description of the available scattering
data and other experimental constraints can be achieved within
the assumption of (broken) SU(3) flavor symmetry. Applications 
of the resulting potentials in bound-state calculations for light hypernuclei 
\cite{YN2019,Le:2020,Le:2021LL,Le:2021XN,Le:2022} 
led to results close to the empirical values. 
In addition, the exploration of neutron-star properties with the strangeness 
$S=-1$ interaction indicate the potential to resolve the so-called hyperon 
puzzle \cite{Chatterjee:2016}, when combined with consistently derived 
$\La NN$ and $\Si NN$ three-body forces \cite{Haidenbauer:2016NS,Gerstung:2020}.
Finally, the $BB$ potentials have been successfully tested in the analysis of 
two-particle momentum correlation functions involving strange baryons
\cite{ALICE:2021L,Haidenbauer:2022F,ALICE:2022LX}. 

In this contribution we present some highlights from recent investigations.
In Sect.~2 preliminary results of an extension of the $\La N$-$\Si N$ interaction up 
to next-to-next-to leading order (N$^2$LO) \cite{YN2022} are reported. 
In Sect.~3 a calculation of charge-symmetry breaking (CSB) of the separation energies 
of the $A$=$4$ $\Lambda$-hypernuclei ${^4_\Lambda \rm He}$ and ${^4_\Lambda \rm  H}$
is reviewed \cite{Haidenbauer:2021CSB}.  
Finally, in Sect.~4 selected results involving the $\Xi N$ interaction \cite{YY2019} are 
summarized. Specifically, predictions for the $\Xi^-p$ momentum correlation function are 
provided \cite{Haidenbauer:2022F}.  


\section{$\La N$-$\Si N$ interaction at next-to-next-to-leading order} 

While the description of the $NN$ interaction within chiral EFT has been already 
pushed up to the fifth order \cite{Reinert:2017,Entem:2017}, corresponding applications 
of that framework to the $YN$ interaction are lagging far behind. Here, NLO is presently 
the state of the art \cite{YN2013,YN2019,YY2015,YY2019}. That status is primarily 
a consequence of the unsatisfactory situation with regard to the data base
were practically only 
cross sections are available and primarily for energies near the thresholds. 
In particular, differential observables that would allow to 
fix the LECs in $P$- and/or higher partial waves, which arise in the chiral 
expansion when going to higher order, are rather scarce and of low statistics. 
Only within the last few years the overall circumstances became more promising,
thanks to experiments performed by the E40 Collaboration at the J-PARC
facility. That collaboration has already published differential cross sections
for the $\Si^+ p$ and $\Si^- p$ channels for momenta from $440$~MeV/c
to $850$~MeV \cite{Miwa:2021,Miwa:2021p,Nanamura:2022} 
and corresponding measurements for $\La p$ are in the stage of
preparation, including possibly even spin-dependent observables. 

This development was one of the reasons to extend our study 
of the $\La N$-$\Si N$ interaction to the next order. 
However, there are also several theoretical aspects which make an
extension to N$^2$LO rather interesting. One of them is that in the
Weinberg counting three-baryon forces emerge at this order. Calculations of 
the four-body systems $^4_\La$H and $^4_\La$He for the NLO13 \cite{YN2013} and 
NLO19 \cite{YN2019} potentials based on the Faddeev-Yakubovsky equations
indicate that the experimental separation energies are underestimated
\cite{YN2019}. Thus, there is obviously a need for including 
$\La NN$ and possibly also $\Si NN$ three-body forces. 
Another appealing factor is (in view of the mentioned scarcity of
data) that no new LECs appear at this order. At the same time pertinent
results for $NN$ scattering indicate that there is some improvement in 
the energy dependence of the $S$-waves and, specifically, in several 
$P$-waves once the contributions involving the sub-leading $\pi N$ 
vertices that enter at N$^2$LO are taken into account. 
Certainly, no major improvement can be expected with regard to the 
residual regulator dependence \cite{Epelbaum:2015}. 
In general, a substantial reduction of regulator artifacts can be only 
achieved by going to high order where then the larger number of LECs allows 
one to absorb the regulator dependence more efficiently.
Since our calculation is at low order it is advantageous to keep
such artifacts as small as possible. For that purpose a novel regularization scheme 
proposed and applied in Ref.~\cite{Reinert:2017} seems to be rather promising. 
Here a local regulator is applied to the pion-exchange contributions and only the 
contact terms, being non-local by themselves, are regularized with a non-local
function. (In earlier works the latter has been applied to the whole 
potential \cite{Epelbaum:2005,YN2013,YN2019}.) Accordingly, 
the resulting interactions were called ``Semilocal momentum-space 
regularized (SMS) chiral $NN$ potentials'' \cite{Reinert:2017}. 
A local regulator for pion-exchange contributions leads to a reduction of 
the distortion in the long-range part of the interaction and, thereby, 
facilitates a more rapid convergence with increasing chiral order.  
Of course, this effect cannot be directly quantified in case of $\La N$ 
and $\Si N$ because of the lack of corresponding empirical information.
Nonetheless, given that we want to compare with the new J-PARC data at 
momenta around $500$~MeV/c a reduction of regulator artifacts is 
definitely desirable.

The extension of our $YN$ interaction up to N$^2$LO builds on the SMS
scheme proposed in Ref.~\cite{Reinert:2017}. Details
of the formalism will be reported elsewhere \cite{YN2022}. 
However, for illustration we show how the non-local exponential regulator employed 
in our $YN$ potentials NLO13 \cite{YN2013} and NLO19 \cite{YN2019} 
for the meson-exchange part is replaced by a local regularization, 
\begin{equation}
V^{\rm non-local}_{P} \propto \frac{e^{-\frac{p'^4 + p^4}{\Lambda^4}}}{\vec q^2 + M^2_P} 
\quad \rightarrow \quad 
V^{\rm local}_{P} \propto \frac{e^{-\frac{\vec q^2 + M^2_P}{\Lambda^2}}}{\vec q^2 + M^2_P} \ .
\label{Eq:reg}
\end{equation}
Here $P$ stands for $\pi$, $K$, or $\eta$,
$\vec p$ and $\vec p'$ are the incoming and outgoing center-of-mass momenta 
of the baryons, $\vec q = \vec p' - \vec p$ is the momentum transfer, and  
$\Lambda$ is the cutoff parameter.

\begin{figure}[t]
\centering
\includegraphics[width=0.51\textwidth]{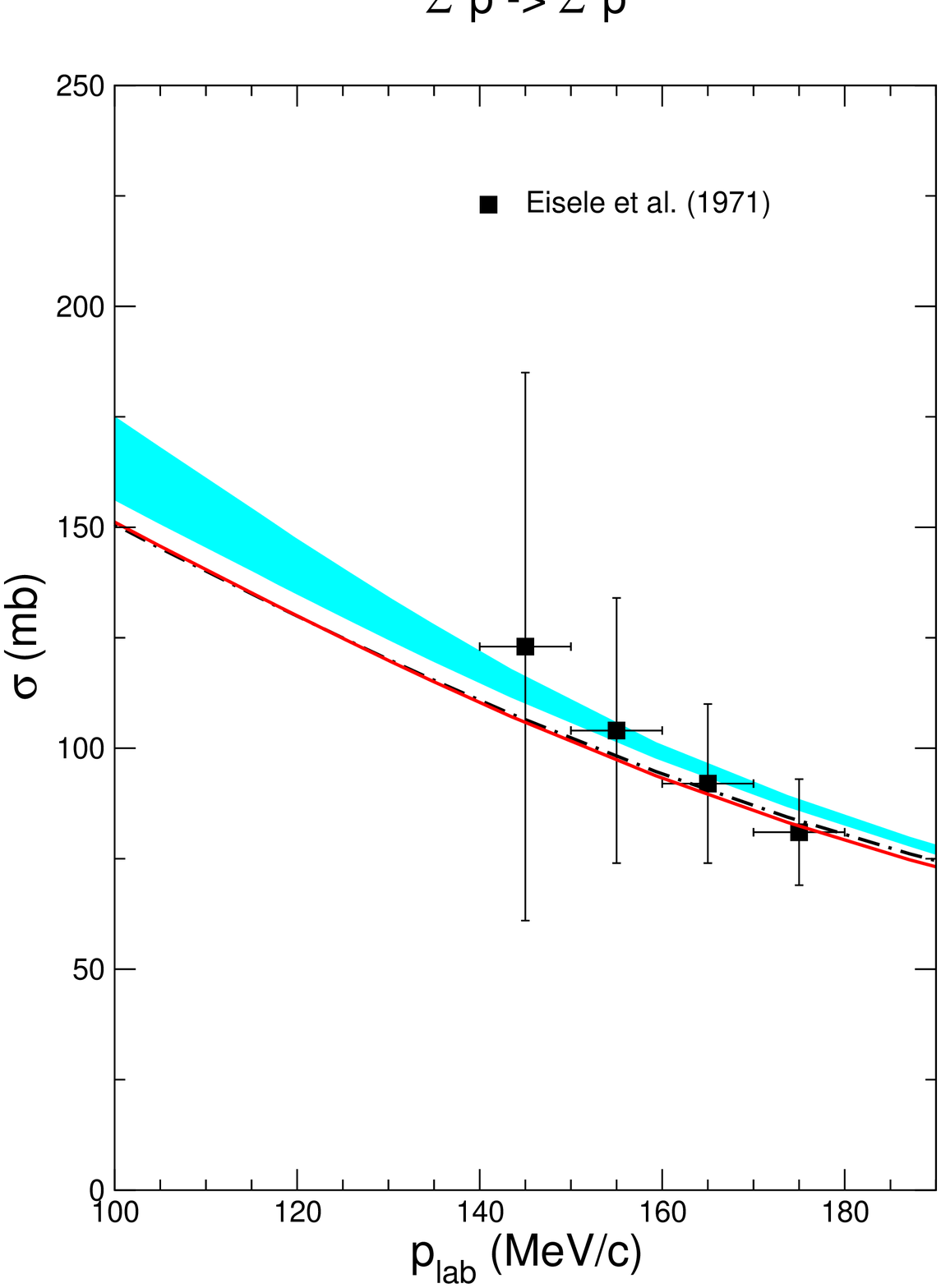}\includegraphics[width=0.51\textwidth]{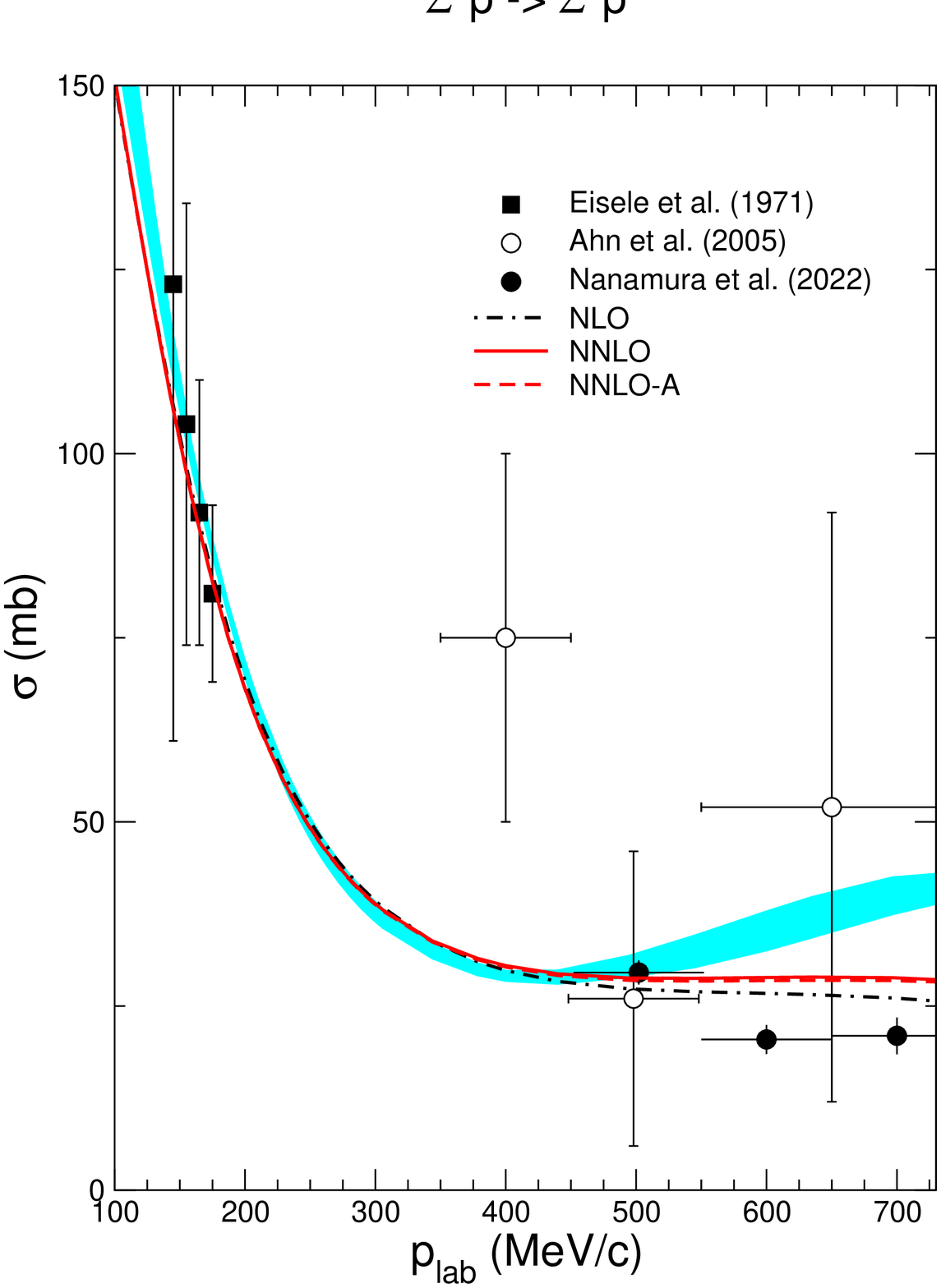}
\vspace*{-1.3cm}
\caption{Cross section for $\Si^+ p$ scattering as a function of $p_{lab}$. Results are shown 
for the SMS NLO (dash-dotted) and N$^2$LO (solid) $YN$ potentials. The dashed line corresponds 
to an alternative fit at N$^2$LO, see text. 
The cyan band is the result for NLO19 \cite{YN2019}. 
Data are from the E40 Collaboration \cite{Nanamura:2022} 
and from Refs.~\cite{Eis71,Ahn05}.  
 \vspace*{-0.7cm}
}
\label{fig:cs5}
\end{figure}
 
In the $NN$ case, where only pion exchanges are taken into account, cutoff values in the 
range $\Lambda = 350-550$ MeV were considered where $\Lambda = 450$ MeV yielded the 
best result \cite{Reinert:2017}. 
The choice of the cutoff mass for the $YN$ interaction is more delicate,
because we want to keep the underlying approximate SU(3) flavor symmetry as well
as the explicit SU(3) breaking in the long-range part of the potential due to 
the mass splitting between the pseudoscalar mesons $\pi$, $K$, and $\eta$. 
Since the kaon mass is around $495$ MeV it seems advisable to use cutoff
masses that are at least $500$~MeV, in order to capture the pertinent
physics. On the other hand, large values, say above $650$~MeV, are questionable 
because, like in $NN$, we want to avoid spurious bound states. 
These considerations suggest that two-meson exchange contributions 
involving a $K$ and/or $\eta$ ($\pi K$, $K K$, etc.) should and can no longer 
be included explicitly but have to be absorbed into the contact terms. Thus, 
contrary to our earlier work \cite{YN2013,YN2019} we allow and expect some SU(3) 
breaking between the LECs in the $\La N$ and $\Si N$ systems. 

In the following we present preliminary results for the cutoff value of $550$~MeV.
We focus on the $\Si N$ channels where new data from J-PARC have become
available. We start the discussion with $\Si^+ p$ scattering which is of 
particular interest for theory. Since the total isospin is $I=3/2$ there is no
coupling to the $\La N$ channel which simplifies the dynamics.  
Moreover, SU(3) symmetry provides strong constraints on several amplitudes.
Specifically, space-spin antisymmetric partial waves ($^1S_0$, $^3P_{0,1,2}$, ...) 
belong all to the $\{27\}$ irrep. of SU(3) symmetry \cite{YN2013,YN2019} and, thus, 
the corresponding interactions would be identical to those in the $NN$ system 
provided that SU(3) symmetry is exactly fulfilled. While
we know that there is a sizable SU(3) breaking in case of the $^1S_0$
partial wave \cite{XX2015}, the amplitudes in the $P$- and higher partial
waves could be much closer to those found for $NN$ scattering.  

\begin{figure}[t]
\centering
\includegraphics[width=0.51\textwidth]{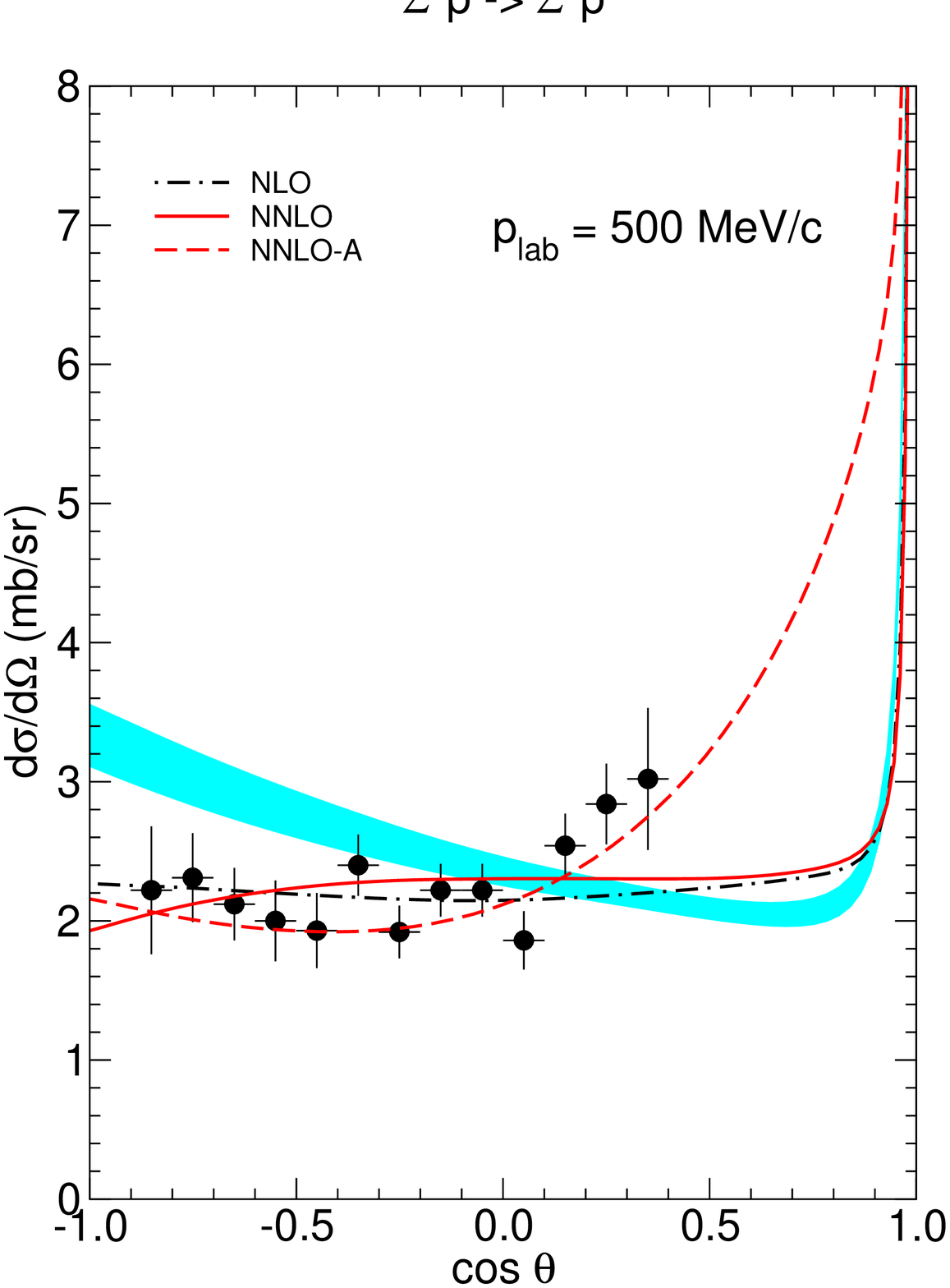}\includegraphics[width=0.51\textwidth]{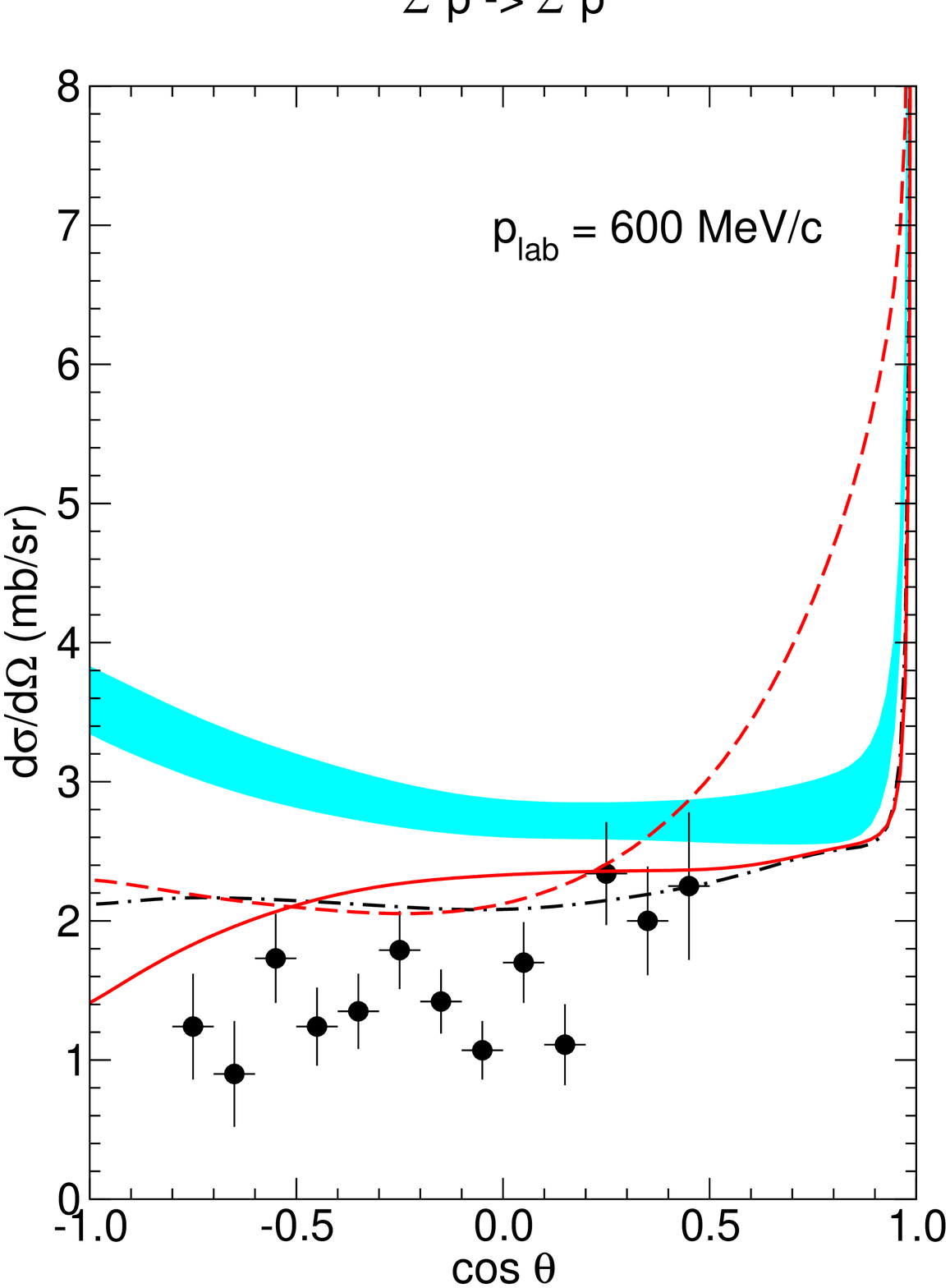}
\vspace*{-1.3cm}
\caption{Differential cross section for $\Si^+ p$. 
Same description of curves as in Fig.~\ref{fig:cs5}. 
Data are from Nanamura et al. \cite{Nanamura:2022}  
(momentum regions $440$-$550$~MeV/c and $550$-$650$~MeV/c).
 \vspace*{-0.7cm}
}
\label{fig:ds5}
\end{figure}

In fitting to the $YN$ data we proceed as before \cite{YN2013,YN2019}, i.e. we consider 
only the set of $36$ data for $\La p$, $\Si^-p$ and $\Si^+p$ scattering at low energies 
for determining the LECs in the $S$ waves. SU(3) symmetry is imposed for the contact
terms at the
initial stage but eventually relaxed for the LO LECs, in line with the 
power counting where SU(3) breaking terms arise from mass insertions in the
chiral Lagrangian at the NLO level \cite{Petschauer:2013}. 
Anyway, as said we do expect some SU(3) breaking in the contacts terms in view of 
the fact that two-meson exchange contributions from $\pi K$, $\pi \eta$, etc. 
are not explicitly included. The achieved $\chi^2$ is comparable to the one found 
for our NLO interactions \cite{YN2013,YN2019}, and typically around 
$16$ for the $36$ data points. The pertinent results for $\Si^+ p$ are
presented in Fig.~\ref{fig:cs5} (left) and compared with data and with
the results obtained from the NLO19 potential. The latter are shown as band, 
representing the cutoff dependence \cite{YN2019}. 

\begin{figure}[t]
\centering
\includegraphics[width=0.51\textwidth]{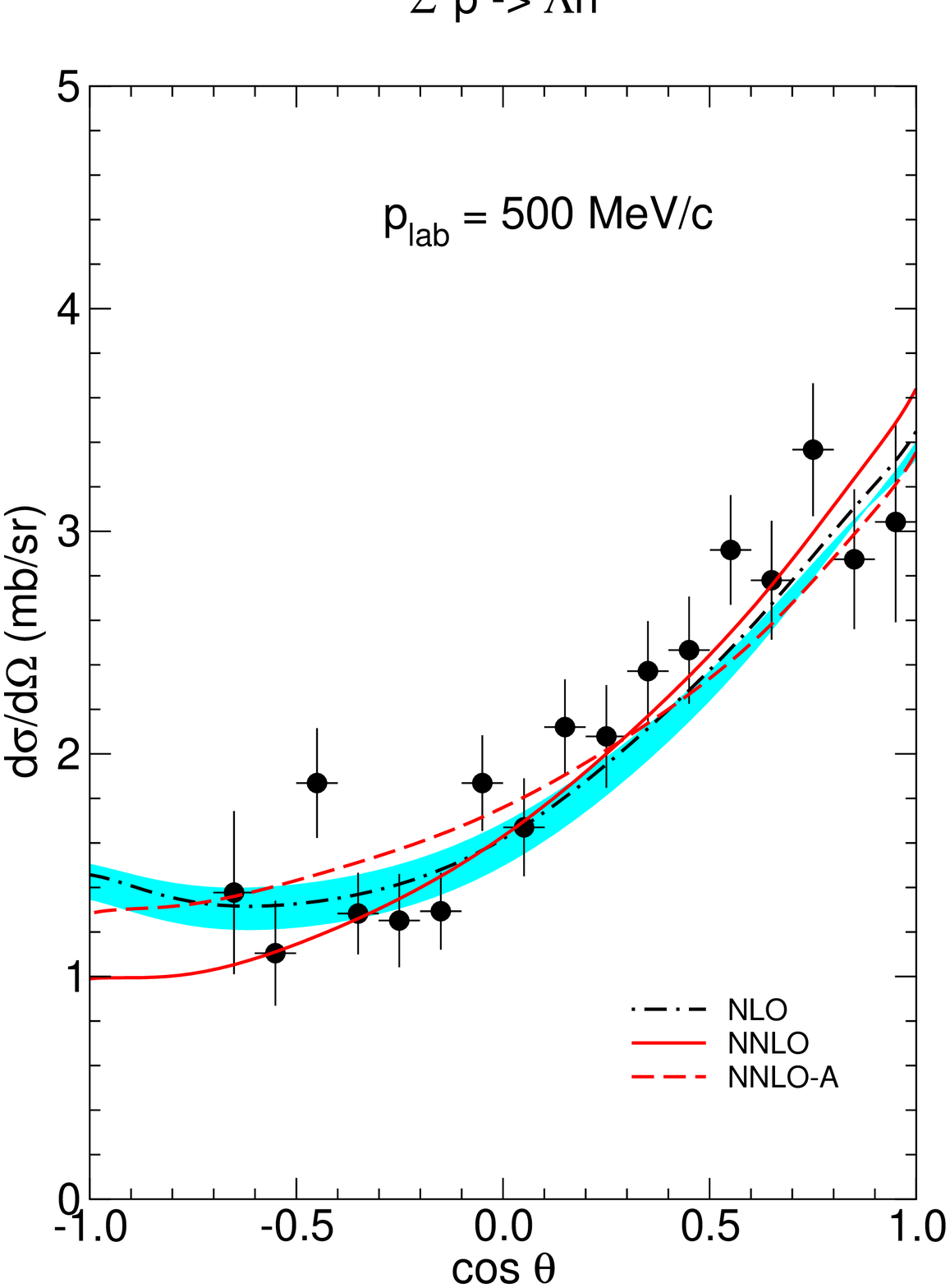}\includegraphics[width=0.51\textwidth]{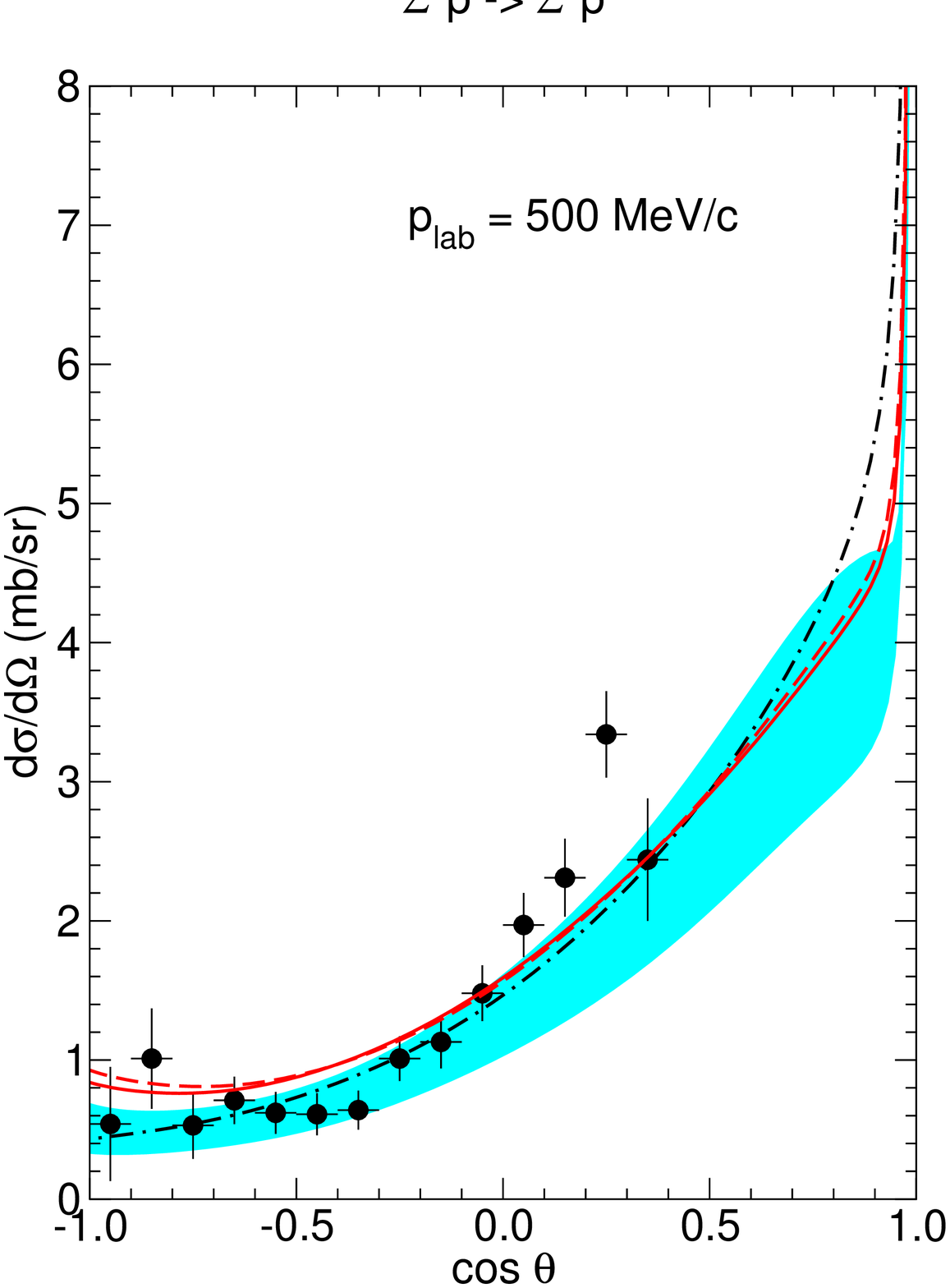}
\vspace*{-1.3cm}
\caption{Differential cross sections for $\Si^- p\to \La n$ and $\Si^- p \to \Si^- p$.
Same description of curves as in Fig.~\ref{fig:cs5}. 
Data are from the E40 Collaboration \cite{Miwa:2021,Miwa:2021p}  
(momentum region $470$-$550$~MeV/c).
\vspace*{-0.7cm}
}
\label{fig:ds2}
\end{figure}

Once the $S$-wave LECs are fixed, 
a fit to the differential cross sections reported by the E40 Collaboration
is performed, starting with the $\Si^+ p$ data for the reasons discussed
above. Interestingly, in the NLO case taking over the LECs in the $^3P_{0,1,2}$ 
partial waves from the corresponding $NN$ potential by Reinert 
et al.~\cite{Reinert:2017}, in accordance with SU(3) symmetry, (and assuming 
the LEC in the $^1P_1$ to be zero) yields already a good description of the 
E40 data taken in the region $440$-$550$~MeV/c, cf. Fig.~\ref{fig:ds5} (left). 
For the N$^2$LO interaction all $P$-wave LECs are fitted to the data. Actually, here we
explore two scenarios, one where the resulting angular distribution is
similar to that obtained for NLO and one which produces an overall more pronounced
angular dependence. The latter is clearly preferred by the available data in that
momentum range. Unfortunately, the data for the next momentum region,
$550$-$650$~MeV/c, suggest an overall somewhat different angular dependence,
see Fig.~\ref{fig:ds5} (right), so that conclusions are difficult to draw
at present. In any case the predictions by NLO19 are clearly at odds with the
data. 

The integrated $\Si^+ p$ cross section over a larger energy range is shown
in Fig.~\ref{fig:cs5} (right). Note that the common angular averaging is performed
here, cf. e.g. Eq.~(24) in Ref.~\cite{YN2013}, for the E40 data (and accordingly 
for the theory results) in order to compensate for the incomplete angular 
coverage. 
Again the NLO19 potential does not reproduce the trend of the data. We believe
that the rise of the cross section for larger $p_{lab}$ could be an artifact
of the employed regulator. Anyway, $p_{lab}=600$~MeV/c corresponds
to a laboratory energy of $T_{lab}\approx 150$~MeV/c so that we are certainly
in a region where NLO and possibly even N$^2$LO cannot be expected to be 
quantitatively reliable. Moreover, one should keep in mind that the 
$\La N\pi$ channel opens around that energy. 

Results for the differential cross sections in the reactions $\Si^-p \to \Si^- p$ and
$\Si^-p \to \La n$ are presented in Fig.~\ref{fig:ds2}. In this case there is good
agreement with the E40 data for all considered $YN$ potentials.

\section{Charge symmetry breaking in $A$=$4$ $\Lambda$-hypernuclei}

In Ref.~\cite{Haidenbauer:2021CSB} we have studied effects from CSB in the $YN$ 
interaction bases on the potentials NLO13 and NLO19.  
Specifically, we have utilized the experimentally known difference 
of the $\Lambda$ separation energies
in the mirror nuclei ${^4_\Lambda \rm He}$ and ${^4_\Lambda \rm  H}$
to constrain the $\Lambda$-neutron interaction.
The CSB part of the $\La N$ potential at NL{\O} \cite{Epelbaum:2005} 
is given by \cite{Haidenbauer:2021CSB} 
\begin{eqnarray}
V^{CSB}_{\La N\to \La N}
&=& \Bigg[ -f^{(\La-\Si^0)}_{\La\La\pi} f_{NN\pi}\frac{\left(\mbox{\boldmath $\sigma$}_1\cdot{\bf q}\right)
\left(\mbox{\boldmath $\sigma$}_2\cdot{\bf q}\right)}{{\bf q}^2+M^2_{\pi^0}} \nonumber\\
&-&f^{(\eta-\pi^0)}_{\La\La\pi} f_{NN\pi}\left(\mbox{\boldmath $\sigma$}_1\cdot{\bf q}\right)
\left(\mbox{\boldmath $\sigma$}_2\cdot{\bf q}\right)
\left( \frac{1}{{\bf q}^2+M^2_{\pi^0}}-\frac{1}{{\bf q}^2+M^2_{\eta}} \right) \nonumber\\
&+&\frac{1}{4}(1-\mbox{\boldmath $\sigma$}_1\cdot\mbox{\boldmath $\sigma$}_2)\, C^{CSB}_{^1S_0}  
\,+\,\frac{1}{4}(3+\mbox{\boldmath $\sigma$}_1\cdot\mbox{\boldmath $\sigma$}_2)\, C^{CSB}_{^3S_1}  
\Bigg] \  \tau_N \ . 
\label{VCSB}
\end{eqnarray}
The CSB contributions arise from a non-zero $\La\La\pi$
coupling constant which is estimated from
$\Lambda-\Sigma^0$ ($f^{(\La-\Si^0)}_{\La\La\pi}$) and $\eta-\pi^0$ 
($f^{(\eta-\pi^0)}_{\La\La\pi}$) mixing \cite{Dalitz:1964}, respectively, 
and from two contact terms, $C^{CSB}_{^1S_0}$ and $C^{CSB}_{^3S_1}$, 
that represent short-ranged CSB forces. 
In addition, there is a small contribution due to the mass difference between
$K^{\pm}$ and $K^0$ \cite{Haidenbauer:2021CSB}.  
$\tau_p = 1$ and $\tau_n =-1$.

In order to fix the LECs $C^{CSB}_{^1S_0}$ and $C^{CSB}_{^3S_1}$ in Eq.~(\ref{VCSB}) 
the observed CSB splittings for the $A=4$ hypernuclei, 
defined in the usual way in terms of the separation energies,
\begin{equation}
\Delta E(0^+) = E^{0^+}_{\La}({^4_\Lambda \rm He})
-E^{0^+}_{\La}({^4_\Lambda \rm H}), 
\qquad
\Delta E(1^+) = E^{1^+}_{\La}({^4_\Lambda \rm He})
-E^{1^+}_{\La}({^4_\Lambda \rm H}) \ , 
\end{equation}
is considered.
We aim at a reproduction of the present experimental situation, established by 
the recent measurements of the ${^4_\Lambda \rm H}$ $0^+$ state in Mainz \cite{Schulz:2016}
and the one of the ${^4_\Lambda \rm He}$ $1^+$-$0^+$ splitting at J-PARC \cite{Yamamoto:2015}, 
which implies $\Delta E(0^+) = 233\pm 92$~keV and $\Delta E(1^+) = -83\pm 94$~keV.

\begin{table}
\caption{$\La p$ and $\La n$ scattering lengths (in fm) in the singlet and triplet $S$ waves for the
NLO19 potential, for cutoffs of $500$-$650$~MeV. In addition, the resulting level splittings for the
$A=4$ mirror nuclei ${^4_\Lambda \rm H}$ and ${^4_\Lambda \rm He}$ (in keV) are listed. 
}
\renewcommand{\arraystretch}{1.3}
\centering
\begin{tabular}{|c||cc|cc||cc|}
\hline
            & $a^{\La p}_{s}$ & $a^{\La n}_{s}$ &
$a^{\La p}_{t}$ & $a^{\La n}_{t}$ &
$\Delta E (0^+)$ & $\Delta E (1^+)$ \\
\hline
\hline
{NLO19}(500) &-2.649 & -3.202 &-1.580 &-1.467 & 249 & -75 \\
{NLO19}(550) &-2.640 & -3.205 &-1.524 &-1.407 & 252 & -72 \\
{NLO19}(600) &-2.632 & -3.227 &-1.473 &-1.362 & 243 & -67 \\
{NLO19}(650) &-2.620 & -3.225 &-1.464 &-1.365 & 250 & -69 \\
\hline
\end{tabular}
 \vspace*{-0.3cm}
\label{tab:CSB} 
\renewcommand{\arraystretch}{1.0}
\end{table}
Our results for NLO19 are summarized in Table~\ref{tab:CSB}. 
One can see that the reproduction of the splittings 
$\Delta E(0^+)$ and $\Delta E(1^+)$
requires a sizable difference between the strength of the
$\La p$ and $\La n$ interactions in the $^1S_0$ state.
The modifications in the $^3S_1$ partial wave
are much smaller and the effect goes also in
the opposite direction, i.e. while for $^1S_0$
the $\La p$ interaction is found to be
noticeably less attractive than $\La n$,
in case of $^3S_1$ it is slightly more attractive.
In terms of the pertinent scattering lengths we
predict for $\Delta a^{CSB} = a_{\La p} - a_{\La n}$
a value of $0.62\pm 0.08$~fm for the $^1S_0$ partial wave
and $-0.10\pm 0.02$~fm for $^3S_1$.
An investigation of CSB in $A=7,8$ $\La$-hypernuclei is in preparation \cite{Le:2022CSB}. 


\begin{figure}
\begin{center}
\includegraphics[height=4.3cm,angle=-90,keepaspectratio]{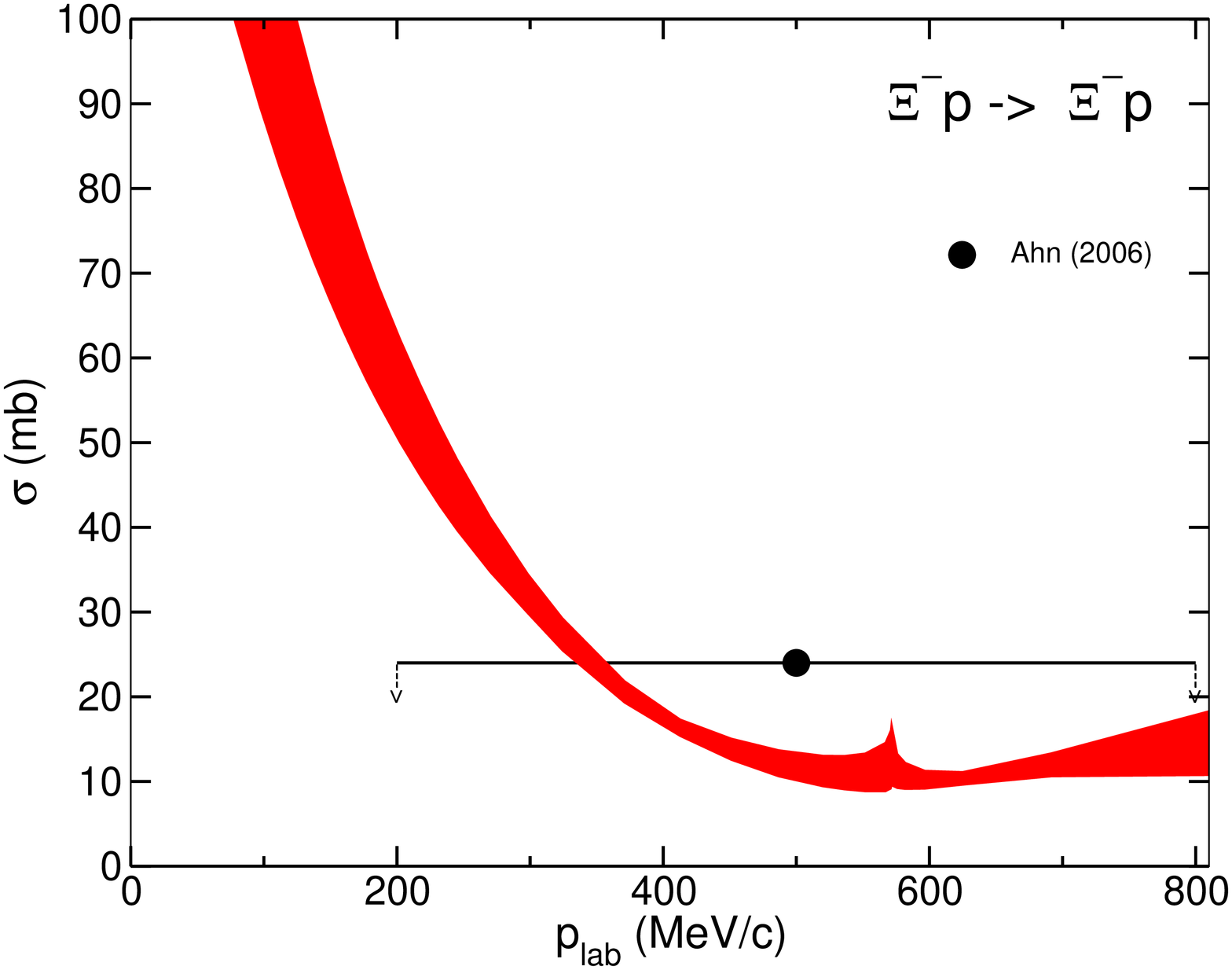}\includegraphics[height=4.3cm,angle=-90,keepaspectratio]{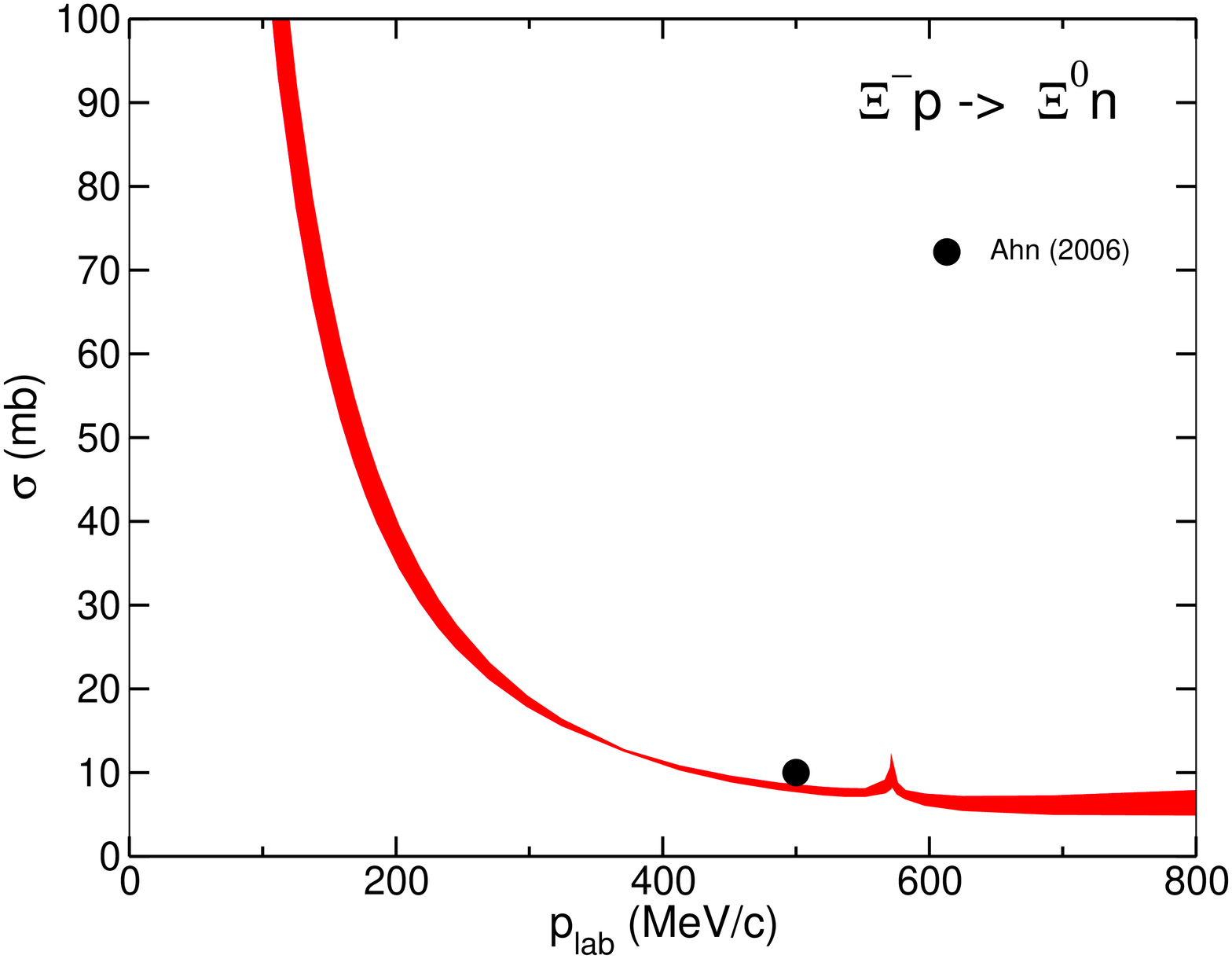}
\includegraphics[height=4.3cm,angle=-90,keepaspectratio]{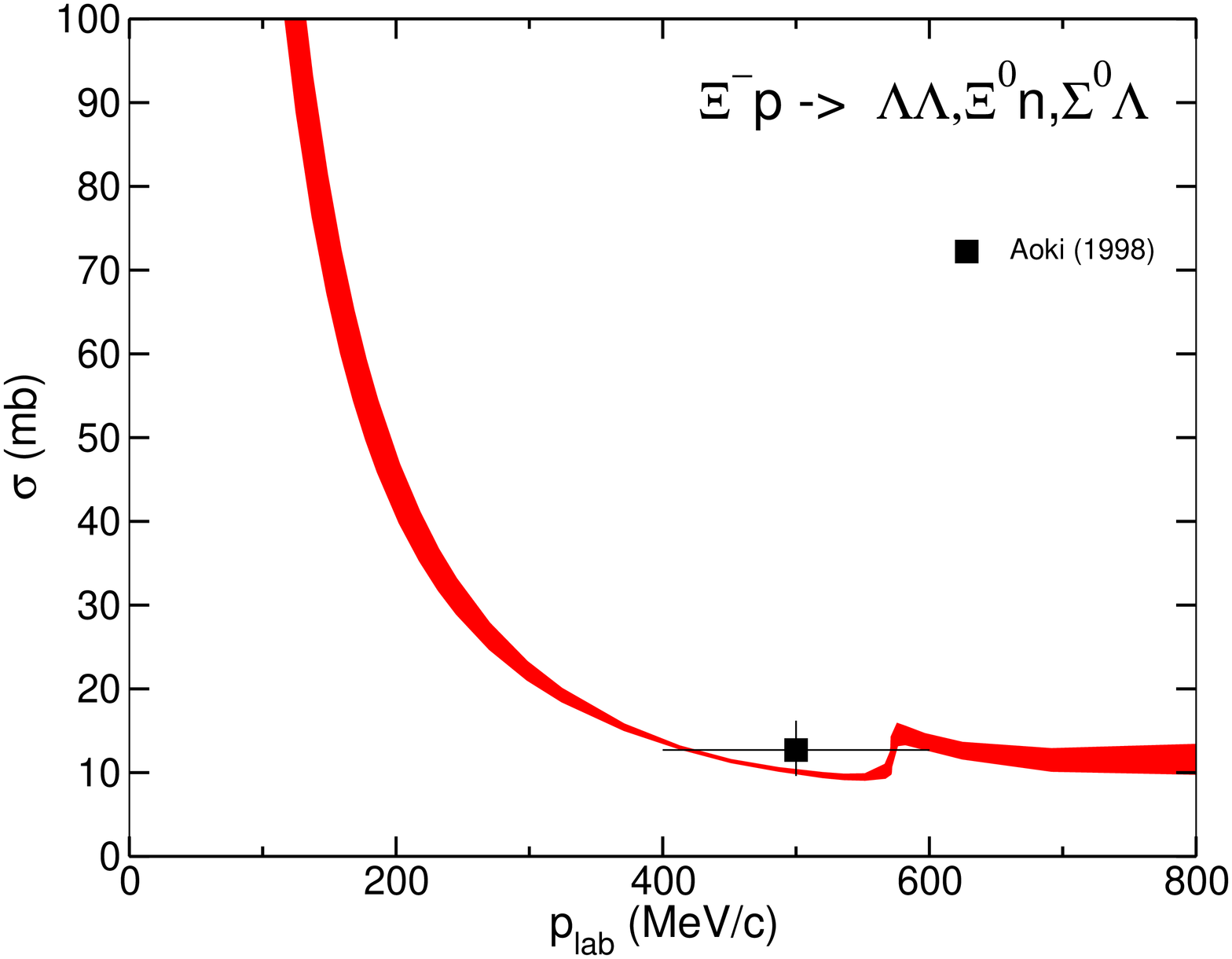}
\end{center}
\vspace*{-0.3cm}
\caption{Results for the $\Xi N$ cross sections of our NLO potential \cite{YY2019}. 
Data are from Refs.~\cite{Ahn:2006,Aoki:1998}. 
\vspace*{-0.6cm}
} 
\label{fig:xmpX}
\end{figure}

\section{Aspects of the $\Xi N$ interaction}

Chiral potentials up to NLO for the $BB$ interaction in the strangeness $S=-2$ system 
$\Xi N$ have been already established by us in 2016 \cite{YY2015}. 
Thereby constraints from the $\La\La$ scattering length in the $^1S_0$ state
together with experimental upper bounds on the cross sections for $\Xi N$
scattering and for the transition $\Xi N \to \La\La$ have been exploited.
This allowed us to fix the additional LECs that arise in the $\{1\}$
irreducible representation of SU(3) \cite{YY2015}. Furthermore,
the consideration of those empirical constraints necessitated to add
SU(3) symmetry breaking contact terms in other irreps
($\{27\}$, $\{10\}$, $\{10^*\}$, $\{8_s\}$, $\{8_a\}$),
with regard to those determined from the $\La N$ and $\Si N$ data.
This is anyway expected and fully in line with the power counting of
SU(3) chiral EFT, as already mentioned in Sect.~2. 
In 2019 a modified version has been suggested \cite{YY2019} which is
more attractive in the $^3S_1$ partial wave with isospin $I=1$.
That potential yields a moderately attractive (in-medium) $\Xi$-nuclear interaction
\cite{YY2019,Kohno:2019} and supports the existence of bound $\Xi$-hypernuclei \cite{Le:2021XN},
in accord with experimental evidence \cite{Nakazawa:2015,Yoshimoto:2021}.
The interactions in the ($I=0,1$) $^1S_0$ partial waves are the same in the 
two versions.

\begin{figure}[htbp]
\centering
\includegraphics[width=0.39\textwidth,angle=-90]{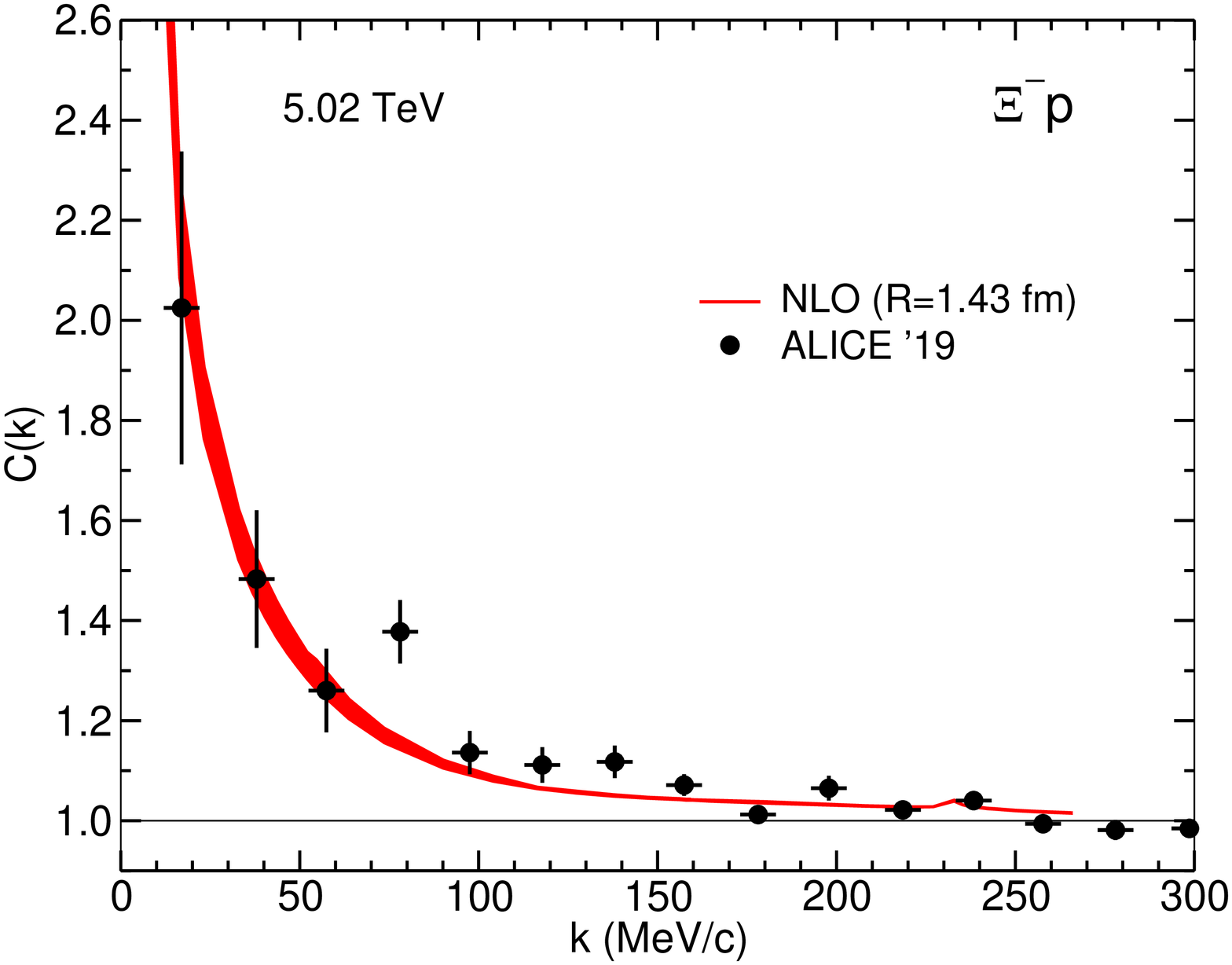}\includegraphics[width=0.39\textwidth,angle=-90]{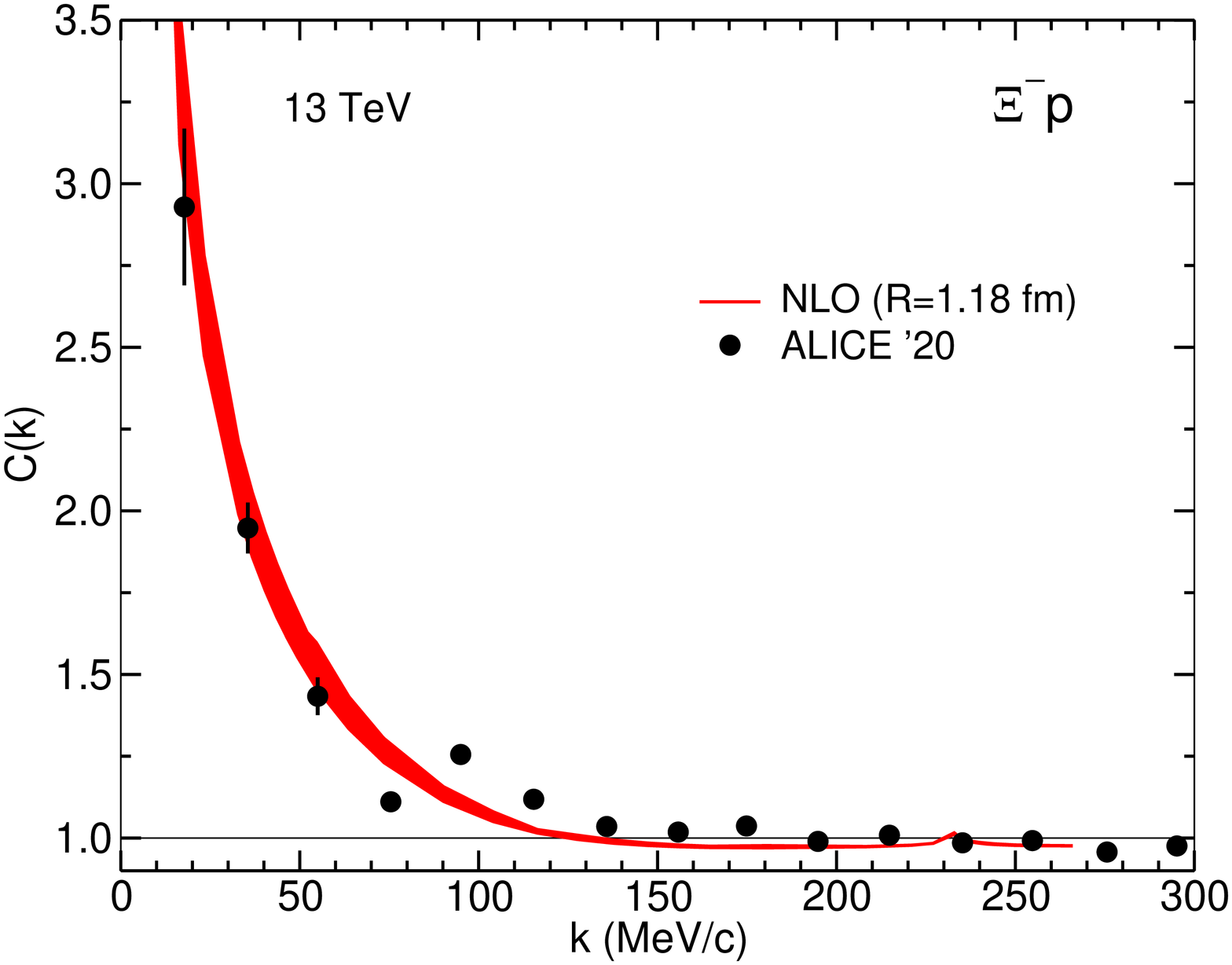}
\vspace*{-0.2cm}
\caption{Predictions for the $\Xi^-p$ two-particle momentum correlation function $C(k)$ 
of our NLO potential \cite{YY2019}. 
Data are from the ALICE Collaboration \cite{ALICE:2019,ALICE:2020}.
\vspace*{-0.6cm}
}
\label{fig:xmp}
\end{figure}

Results for the $\Xi N$ cross sections based on the $\Xi N$ potential from 2019
are presented in Fig.~\ref{fig:xmpX}. 
An interesting and independent test for the $\Xi N$ interaction is provided by 
two-particle momentum correlations. For $\Xi^- p$, correlation functions 
have been measured recently by the ALICE Collaboration in $p$-Pb collisions 
at $5.02$ TeV \cite{ALICE:2019} and in $pp$ collisions at $13$ TeV \cite{ALICE:2020}. 
There are also new and still preliminary results from Au+Au collisions at $200$ GeV by 
the STAR Collaboration \cite{STAR:2021}.
In Fig.~\ref{fig:xmp} we present predictions for $C(k)$ for the $S=-2$ interaction
from 2019. The bands reflect the residual cutoff dependence \cite{YY2019}. 
Details on the evaluation of such correlation functions can 
be found, e.g., in Refs.~\cite{Haidenbauer:2018,Kamiya:2021}. 
Clearly, the correlation functions, evaluated for the source radii
$R$ taken from the corresponding $pp$ fits by ALICE~\cite{ALICE:2019S}
($1.43$~fm for $5.02$~TeV and $1.18$~fm for $13$~TeV), agree nicely with
the measurements. 
For a more thorough discussion on the choice of $R$ and of the other parameters
that enter into the calculation, see \cite{Haidenbauer:2022F,Kamiya:2021}. 

{\small
{\bf Acknowledgements:}
Work supported by the European Research Council (ERC) under the European
Union's Horizon 2020 research and innovation programme 
(grant no.~101018170, EXOTIC), and by the DFG and the NSFC through
funds provided to the Sino-German CRC 110 ``Symmetries and
the Emergence of Structure in QCD'' (DFG grant. no.~TRR~110). 
}

\end{document}